\documentclass[runningheads]{llncs}
\usepackage{etex}
\usepackage[latin1]{inputenc}
\usepackage{mathptmx}
\usepackage{latexsym}
\usepackage{amsfonts,amssymb,latexsym}
\usepackage{amsmath}
\usepackage{enumitem}
\usepackage{array,tabularx}
\usepackage{float}
\usepackage{graphicx}
\usepackage{tikz,pgfplots}
\usepackage{wrapfig}
\usepackage{algorithm}
\usepackage[noend]{algpseudocode}
\usepackage[all]{xy}
\usepackage{verbatim} 
\usepackage{url} 
\usepackage[T1]{fontenc}
\usepackage{multirow}
\usepackage{stackrel}

\usepackage{geometry} 
\newcommand{\blacktrianglelefteq}{\stackrel{\blacktriangleleft}{\_}}
\newcommand{\blacktrianglelefteqvar}{\stackrel[]{\blacktriangleleft}{\_}^v}
\newcommand{\trianglelefteqgd}{\trianglelefteq^{gd}}
\newcommand{\trianglelefteqrogd}{\trianglelefteq^{rogd}}

\newcommand{\trianglelefteqdestrat}{\trianglelefteq^{sml}}
\newcommand{\trianglelefteqdeml}{\trianglelefteq^{ml}}

\newcommand{\Symbols}{\Sigma}

\newcommand{\domain}[1]{\mathit{Dom}(#1)}
\newcommand{\range}[1]{\intrvar{#1}}
\newcommand{\intrvar}[1]{\mathit{Ran}(#1)}

\let\oldAA\AA
\renewcommand{\AA}{\text{\normalfont\oldAA}}

\def\defemb#1#2{\expandafter\def\csname #1\endcsname
	{\relax\ifmmode #2\else\hbox{$#2$}\fi}}

\def\caE{\mathcal{E}}

\def\cN{\mathcal{N}}
\def\cV{\mathcal{V}}

\def\cP{\mathcal{P}}
\def\cR{\mathcal{R}}

\long\def\comment#1{}

\newcommand{\idsubst}{\textit{id}}
\newcommand{\var}{{\cV}ar}

\def\defemb#1#2{\expandafter\def\csname #1\endcsname
	{\relax\ifmmode #2\else\hbox{$#2$}\fi}}

\def\ll{[\![}
\def\rr{]\!]}
\def\Den#1{\relax\ifmmode \ll #1\rr \else\hbox{$\ll #1\rr$}\fi}

\let\|=\mid

\def\#{\hat{~}}

\defemb{implies}{\quad\rightarrow\quad}
\defemb{Implies}{\quad\Rightarrow\quad}

\defemb{cA}{{\cal A}}
\defemb{cB}{{\cal B}}
\defemb{cC}{{\cal C}}
\defemb{cD}{{\cal D}}
\defemb{cE}{{\cal E}}
\defemb{cF}{{\cal F}}
\defemb{cG}{{\cal G}}
\defemb{cH}{{\cal H}}
\defemb{cI}{{\cal I}}
\defemb{cJ}{{\cal J}}
\defemb{cL}{{\cal L}}
\defemb{cP}{{\cal P}}
\defemb{cR}{{\cal R}}
\defemb{cS}{{\cal S}}
\defemb{cT}{{\cal T}}
\defemb{caT}{{\cal T}}
\defemb{cU}{{\cal U}}
\defemb{cV}{{\cal V}}
\defemb{cX}{{\cal X}}
\defemb{caX}{{\cal X}}
\defemb{cZ}{{\cal Z}}

\long\def\comment#1{}

\defemb{cR}{{\cal R}}
\defemb{cE}{{\cal E}}
\defemb{cN}{{\cal N}}
\defemb{cV}{{\cal V}}
\defemb{cP}{{\cal P}}
\defemb{cX}{{\cal X}}
\defemb{cU}{{\cal U}}

\newcommand{\Xc}{{\mathcal{X}}}

\newcommand{\Tc}{{\mathcal{T}}}
\newcommand{\Var}{{\mathcal V}ar}
\newcommand{\pr}[1]{\mbox{$\tt #1$}}

\newcommand{\toppos}{\mbox{\footnotesize$\Lambda$}}
\newcommand{\pos}{{\cP}os}
\newcommand{\fpos}{{\cN}{\cV}{\cP}os}

\newcommand{\rewrite}[1]{\rightarrow_{#1}}
\newcommand{\rewrites}[1]{\rightarrow^*_{#1}}

\newcommand{\TermsOn}[5]{{\caT^{#4}_{#1}(#2)}_{#3}^{#5}}
\newcommand{\Terms}{\TermsOn{\Symbols}{\Variables}{}{}{}}
\newcommand{\Variables}{\caX}
\newcommand{\funocc}[1]{\mathit{Pos}_{\Symbols}(#1)}
\newcommand{\subterm}[2]{#1|_{#2}} 
\newcommand{\congr}[1]{=_{\protect #1}}
\newcommand{\replace}[3]{#1[#3]_{#2}}
\newcommand{\norm}[1]{{!_ {#1}}}
\newcommand{\TermsS}[1]{\TermsOn{\Symbols}{\Variables}{\sort{#1}}{}{}}
\newcommand{\sort}[1]{\ensuremath{\mathsf{#1}}}

\newcommand\bcmdtab{\noindent\bgroup\tabcolsep=0pt%
  \begin{tabular}{@{}p{10pc}@{}p{20pc}@{}}}
\newcommand\ecmdtab{\end{tabular}\egroup}

\usepackage{array}
\newcolumntype{C}[1]{>{\centering\let\newline\\\arraybackslash\hspace{0pt}}m{#1}}

\begin{document}
  \title{Homeomorphic Embedding modulo Combinations of Associativity and Commutativity Axioms
  \thanks{This work has been
        		partially supported by the EU (FEDER) and the Spanish
       			MINECO under grant TIN 2015-69175-C4-1-R,
				and
		        by Generalitat Valenciana under grant PROMETEOII/2015/013.
				Jose Meseguer was partially supported by NRL under contract number N00173-17-1-G002.
		        Angel Cuenca-Ortega has been
        		supported by the SENESCYT, Ecuador (scholarship program 2013)
        	}
	}

  \author{
	  Mar\'{\i}a Alpuente\inst{1}
	  \and
	  Angel Cuenca-Ortega\inst{1,3}
	  \and
	  Santiago Escobar\inst{1}
	  \and
	  Jos\'e Meseguer\inst{2}
	}

\institute{
	DSIC-ELP, Universitat Polit\`ecnica de Val\`encia, Spain.
	\email{\{alpuente,acuenca,sescobar\}@dsic.upv.es}
	\and
	University of Illinois at Urbana-Champaign, USA.
	\email{meseguer@illinois.edu}
	\and
	Universidad de Guayaquil, Ecuador.
	\email{angel.cuencao@ug.edu.ec}
}

\maketitle

\begin{abstract}
		The  Homeomorphic Embedding  relation has been amply used for    defining termination criteria of symbolic methods for program analysis, transformation, and verification.  However, homeomorphic embedding has never been  investigated in the context of  
		order-sorted rewrite  
		theories  that support
		 symbolic execution  methods \emph{modulo} equational axioms. 
		This paper generalizes the symbolic homeomorphic embedding relation to order--sorted 
		 rewrite theories that may contain various combinations of associativity and/or commutativity  axioms for different binary operators. 
		We systematically measure the   performance of increasingly efficient formulations of the  homeomorphic embedding relation modulo associativity and commutativity axioms. From our experimental results, we conclude that our most efficient version indeed pays off in  practice.
\end{abstract}

\pagestyle{plain}

\section{Introduction}
\textit{Homeomorphic Embedding}  is a control mechanism that   is commonly used to ensure termination of 
symbolic methods 
and program  optimization techniques.
Homeomorphic embedding is a  structural
preorder relation under which a term $t'$ is greater than (i.e., it embeds) another term $t$ represented by $t \trianglelefteq t'$ if $t$ can be obtained from $t'$ by 
deleting some symbols of $t'$. 
For instance, $v=s(0+s(X))*s(X+Y)$ embeds $u=s(X)*s(Y)$.
The usefulness of   homeomorphic embedding for ensuring termination
is given by the following well-known property of well-quasi-orderings:
given a finite signature, for every 
infinite sequence of terms 
$t_1, t_2, \ldots,$ there exist  $i<j$ such that $t_i \trianglelefteq t_j$.
Therefore, if we iteratively compute 
a sequence $t_1, t_2, \ldots,t_n$,  we can guarantee finiteness of the sequence   by using the embedding as a whistle:
whenever a new expression $t_{n+1}$ is to
be added to the sequence, we first check whether $t_{n+1}$ embeds any
of the expressions that are already in the sequence. 
If that is the case, the computation
must be stopped because the whistle   ($\trianglelefteq$)  signals
 (potential) non-termination. Otherwise, $t_{n+1}$ can be safely added to the sequence and the
computation 
proceeds. 

In \cite{ACEM17}, an order-sorted extension of homeomorphic embedding modulo equational axioms, such as associativity and commutativity,
 was 	defined as a key component  of the symbolic�partial evaluator  Victoria. Unfortunately, the formulation in \cite{ACEM17} was done with a concern 
 for simplicity in mind and degrades the tool performance because the proposed implementation of equational homeomorphic embedding
	did not
	scale well to realistic  problems. This was not unexpected since other equational  problems (such as equational matching, equational unification, 
	or equational least general generalization) are typically much more involved than their corresponding ``syntactic'' counterparts, and achieving efficient implementations 
	has required years of significant investigation effort.
	
	\paragraph{\bf Our contribution.} 
	
 In this paper, we introduce four different  formulations of order-sorted homeomorphic embedding modulo axioms 
in rewrite theories that may contain sorts, subsort polymorphism,  overloading, and rewriting with (conditional) rules and equations modulo a set $B$ of equational axioms,
and we compare their performance. 
We propose an order-sorted, equational homeomorphic embedding formulation $\trianglelefteqdestrat_B$
that runs up to 5 orders of magnitude faster than the original  definition of $\trianglelefteq_B$ in \cite{ACEM17}. 
For this improvement in performance, we take 
advantage of  Maude's powerful capabilities such as  the efficiency of
 deterministic computations with equations versus non-deterministic  computations with rewriting rules,  or the use of non-strict definitions  of the  boolean operators versus more speculative standard boolean definitions 
 \cite{maude-book}. 
 
\paragraph{\bf Plan of the paper.}
	 After some preliminaries in Section \ref{sec:prelim},  Section \ref{sec:lopstr}   recalls the (order-sorted)  homeomorphic equational embedding  relation  
	of \cite{ACEM17}
that extends   the ``syntactically simpler'' homeomorphic embedding    
on nonground terms to the order-sorted case \emph{modulo} equational axioms. 
Section \ref{sec:gd}   provides  two {\em goal-driven} formulations for equational homeomorphic embedding: first, a calculus for embeddability goals
that directly handles the algebraic axioms in the deduction system, and then  a reachability oriented  characterization that  
cuts down the search space by taking advantage of pattern matching modulo associativity and commutativity axioms.
Section \ref{sec:cinco} is concerned with an efficient meta-level   formulation of  equational  homeomorphic  embedding    
that relies on the classical flattening transformation that canonizes terms w.r.t.\ associativity and/or commutativity axioms (for instance, $1 + (2+3)$ gets flattened to $+(1,2,3)$).   An improvement of the algorithm is also achieved by  replacing the classical boolean operators by  short-circuit, strategic versions of these operators.
We provide an experimental performance evaluation of the proposed  formulations showing that we can efficiently deal with realistic embedding problems modulo 
 axioms.  
 	
\section{Preliminaries}\label{sec:prelim}
	Given an \emph{order-sorted signature} $\Sigma$, with a finite poset of sorts $(S, \leq)$, we consider an {$\sort{S}$}-sorted family {$\Xc = \{\Xc_s\}_{\sort{s} \in \sort{S}}$} of disjoint variable sets. {$\Tc_{\Sigma}(\Xc)_\sort{s}$ and ${\Tc_{\Sigma}}_\sort{s}$ denote the sets of terms and ground terms of sorts $\sort{s}$, respectively}. 
	We also write $\Tc_{\Sigma}(\Xc)$ and $\Tc_{\Sigma}$ for the corresponding term algebras.
	In order to simplify the presentation, we often disregard
	sorts when no confusion can arise. 
	
	A \emph{position} $p$ in a term $t$ is represented by a sequence of
	natural numbers ($\toppos$ denotes the empty sequence, i.e., the
	root position).
	Positions are ordered by the \emph{prefix} ordering: $p \leq q$ if
	there exists $w$ such that $p.w =q$.
	Given a term $t$, we let $\pos(t)$ and $\fpos(t)$ respectively denote the set of
	positions and the set of non-variable positions of $t$ (i.e.,
	positions where a variable does not occur). 
	$t|_p$ denotes the \emph{subterm} of $t$ at position
	$p$, and $t[u]_p$ denotes the result of
	\emph{replacing the subterm} $t|_p$ by the term $u$.
	The set of variables occurring in a term $t$ is denoted by $\Var(t)$.
	 
	A \textit{substitution} $\sigma$ is a sorted mapping from a finite
	subset of $\Variables$ to $\Terms$.
	Substitutions are written as 
	$\sigma=\{X_1 \mapsto t_1,\ldots,X_n \mapsto t_n\}$ where
	the domain of $\sigma$ is
	$\domain{\sigma}=\{X_1,\ldots,X_n\}$
	and 
	the set
	of variables introduced by terms $t_1,\ldots,t_n$ is written $\range{\sigma}$.  
	The identity
	substitution is $\idsubst$.  Substitutions are homomorphically extended
	to $\Terms$. 
	The application of a substitution $\sigma$ to a term $t$ is called \emph{an instance} of $t$ and is
	denoted by $t\sigma$. 
	For simplicity, we assume that every substitution is idempotent,
	i.e., $\sigma$ satisfies $\domain{\sigma}\cap\range{\sigma}=\emptyset$.
	Substitution idempotency ensures $(t\sigma)\sigma=t\sigma$.
	The restriction of $\sigma$ to a set of variables
	$V$ is denoted $\subterm{\sigma}{V}$.
	Composition of two substitutions is denoted by $\sigma\sigma'$ so that $t(\sigma\sigma')=(t\sigma)\sigma'$.
	
	A \textit{$\Symbols$-equation} is an unoriented pair $t = t'$, where
	$t,t' \in \TermsS{s}$ for some sort {$\sort{s}\in\sort{S}$}.  Given
	$\Symbols$ and a set $E$ of $\Symbols$-equations, 
	order-sorted
	equational logic induces a congruence relation $\congr{E}$ on terms
	$t,t' \in \Terms$ (see~
	\cite{DBLP:journals/tcs/BouhoulaJM00}). 
	An \emph{equational theory} $(\Symbols,E)$ 
	is a pair with $\Symbols$ being an order-sorted signature and $E$ 
	a set of 
	$\Symbols$-equations. 
	We omit $\Sigma$ when no confusion can arise.
	
	A substitution $\theta$ is more (or equally) general than $\sigma$ modulo $E$, 
	denoted by $\theta \leq_{E} \sigma$, if 
	there is a substitution $\gamma$ such that $\sigma \congr{E} \theta\gamma$, i.e.,\ for all $x \in {\caX}, x\sigma =_E x\theta\gamma$.
	A substitution $\sigma$ is called a renaming if $\sigma=\{X_1 \mapsto Y_1,\ldots,X_n \mapsto Y_n\}$,
	the sorts of $X_i$ and $Y_i$ coincide,
	and variables $Y_1,\ldots,Y_n$ are pairwise distinct. 
	The renaming substitution $\sigma$ is a renaming for expression $E$ if  
	($\Var(E)- \{X_,\ldots,X_n\})\cap\{Y_1,\ldots,Y_n\}=\emptyset$.
	
	An \textit{$E$-unifier} for a $\Symbols$-equation $t = t'$ is a
	substitution $\sigma$ such that $t\sigma \congr{E} t'\sigma$.  
	An $E$-unification algorithm is \textit{complete} if for any equation $t
	= t'$ it generates a complete set of $E$-unifiers, which is defined by the property that the set of all $E$-instances of its elements is exactly the set of all $E$-unifiers.  
	Note that this set
	does not need to be finite. 
	A unification algorithm is said to be
	\textit{finitary} and complete if it always terminates after
	generating a finite and complete set of unifiers.
	
	A \emph{rewrite theory} is a triple {$\cR = (\Sigma,E,R)$}, where $(\Sigma,E)$ 
	is the equational theory  
	modulo that we rewrite  and $R$ is a set of rewrite rules. Rules are of the form $l \to r$ where 
	terms 
	$l,r\in\TermsS{\sort{s}}$ for some sort \sort{s}
	are respectively called the \emph{left-hand side}
	(or \emph{lhs}) and the \emph{right-hand side} (or \emph{rhs}) of the rule 
	and $\Var(r) \subseteq \Var(l)$.  Let $\to\: \subseteq A\times A$ be a binary relation on a set $A$.
	We denote its transitive closure by $\to^+$, and its reflexive and transitive closure by $\to^*$.
	
	We define the {\em one-step rewrite relation} on $\Tc_\Sigma(\Xc)$ for the set of rules $R$ as follows: $t \to_R t'$ iff there is a position $p \in \pos(t)$, a rule $l \to r$ in $R$, and a substitution $\sigma$ such that $t|_p = l\sigma$ and $t' = t[r\sigma]_p$. 
	The relation $\to_{R/E}$ for rewriting modulo $E$ is defined as $=_E \circ \to_R \circ =_E$.
	A term $t$ is called $R/E$-\emph{irreducible} iff there is no term $u$ such that $t \to_{R/E} u$.
	A substitution $\sigma$ is $R/E$-irreducible
	if, for every $x\in\Variables$, $x\sigma$ is 
	$R/E$-irreducible.
	We say that the relation $\rewrite{R/E}$ is
	\emph{terminating} if there is no infinite sequence $t_1 \rewrite{R/E}
	t_2 \rewrite{R/E} \cdots t_n \rewrite{R/E} t_{n+1} \cdots$.  We say that the relation 
	$\rewrite{R/E}$
	is 
	\emph{confluent} 
	if, 
	whenever $t \rewrites{R/E} t'$ and $t
	\rewrites{R/E} t''$, 
	there exists a term $t'''$ such that $t'
	\rewrites{R/E} t'''$ and $t'' \rewrites{R/E} t'''$.  
	We say that $\rewrite{R/E}$
	is \emph{convergent} if it is confluent and terminating.
	An order-sorted rewrite theory $(\Symbols,E,R)$
	is convergent (resp. terminating, confluent) 
	if the relation
	$\rewrite{R/E}$ is 
	convergent (resp. terminating, confluent).
	In a 
	confluent, terminating,
	order-sorted rewrite theory, 
	for each term $t\in\Terms$, there is a unique (up to $E$-equivalence) 
	$R/E$-irreducible term $t'$ that can be 
	obtained by rewriting $t$ to  
	$R/E$-irreducible or {\em normal} form, 
	which is denoted
	by $t \rewrite{R/E}^! t'$, or
	$t{\norm{R/E}}$ when $t'$ is not relevant. 
	
	Since $E$-congruence classes can be infinite,
	$\rewrite{R/E}$-reducibility is undecidable in general.  Therefore, $R/E$-rewriting is usually implemented 
	by $R{,}E$-rewriting.
	We define the relation $\rewrite{R,E}$ on $\Terms$ by 
	$t \rewrite{p,R,E} t'$ (or simply $t \rewrite{R,E} t'$)
	iff there is a non-variable position $p \in \funocc{t}$, 
	a rule $l \to r$ in $R$,
	and a substitution $\sigma$ such that 
	$\subterm{t}{p} \congr{E} l\sigma$
	and $t' = \replace{t}{p}{r\sigma}$.  
	To ensure
	completeness of $R{,}E$-rewriting w.r.t.\ $R/E$-rewriting, we require
	\emph{strict coherence},  ensuring that  $=_E$ is a bisimulation for $R,E$-rewriting \cite{meseguer-coherence}:
	for any $\Sigma$-terms $u,u',v$
	if $u=_{E}u'$ and $u\rightarrow_{R,E}v$, then there exists a term $v'$
	such that $u'\rightarrow_{R,E}v'$ and $v=_{E}v'$. 
	Note that, assuming $E$-matching is decidable, $\rewrite{R,E}$ is decidable
	and notions such as confluence, termination, irreducible term, 
	and
	normalized substitution, 
	are defined for $\rewrite{R,E}$  straightforwardly \cite{meseguer-coherence}.
	It is worth noting that Maude automatically provides  $B$-coherence completion for rules and equations \cite{meseguer-coherence}.
	
	Algebraic structures often involve axioms like associativity (A) and/or commutativity (C)  
	of function symbols, which cannot be handled by ordinary term rewriting but instead are  handled implicitly by working with congruence classes of terms. This is why often an  
		equational theory $E$ is decomposed into a disjoint union  $E= E_{0} \uplus B$,  where the set $E_{0}$ consists of (conditional) equations 
		that are implicitly oriented from left to right as rewrite rules (and operationally used as simplification rules), and  $B$ is a set  of algebraic axioms (which are implicitly expressed  in Maude as  attributes of their corresponding operator using the \pr{assoc} and  \pr{comm}  
		keywords) that are only used for $B$-matching.
	
	We formalize the notion of \emph{decomposition} of an equational theory $(\Symbols,E_0\uplus B)$ into
	a (well-behaved) rewrite theory $(\Symbols,B,\overrightarrow{E_0})$ that  satisfies
	all of the conditions we need, where equations in $E_0$ are {\em explicitly oriented} from left to right as
	$\overrightarrow{E_0}=\{t \to t' \mid t = t'\in E_0\}$.  In a decomposition,  the oriented equations in $\overrightarrow{E_0}$ are used as 
	simplification rules,  and the  algebraic axioms of $B$ are used for $B$-matching (and are never used for rewriting).
	
	\begin{definition}[Decomposition {\rm 
	\cite{DBLP:journals/tcs/BouhoulaJM00}}]
		Let $(\Symbols,E)$ be an order-sorted equational theory.
		We call $(\Symbols,B,\overrightarrow{E_0})$  a \emph{decomposition}
		of $(\Symbols, E)$ if $E = E_0 \uplus B$ and
		$(\Symbols,B,\overrightarrow{E_0})$ is an order-sorted rewrite theory satisfying the following properties:
		
		\begin{enumerate}
			\item\label{ppty1} $B$ is 
			\emph{regular},
			i.e.,\ for each $t = t'$ in
			$B$, we have $\var{(t)} = \var{(t')}$, and
			\emph{linear},
			i.e.,\ for each $t = t'$ in
			$B$, each variable occurs only once in $t$ and in $t'$.
			\item
			$B$ is 
			\emph{sort-preserving}, i.e.,\
			for each  $t = t'$ in
			$B$, 
			sort \sort{s}, and 
			substitution $\sigma$, we have $t \sigma \in
			\TermsS{s}$ iff $t' \sigma \in \TermsS{s}$; furthermore, 
			for each  $t = t'$ in $B$, all variables in $\var{(t)} \cup \var{(t')}$
			have a top\footnote{The poset $(\sort{S},\leq)$ of sorts for $\Symbols$
				is partitioned into equivalence classes (called \emph{connected components})
				by the equivalence relation $(\leq \cup \geq)^+$.
				We assume that each connected component
				$[\sort{s}]$  has a \emph{top sort element} under $\leq$,
				denoted $\top_{[\sort{s}]}$. 
				This involves no real loss of generality,
				since if $[\sort{s}]$ lacks a top sort, it can easily be  added.} sort.
			\item\label{ppty2} $B$ has a finitary and complete matching algorithm so that $B$-matching is decidable\footnote{The definition in \cite{DBLP:journals/entcs/EscobarMS09} requires that $B$-unification is decidable.}. 
			\item\label{ppty5} The rewrite rules in $\overrightarrow{E_0}$  
			are \emph{convergent},
			i.e. confluent, terminating, and strictly coherent modulo $B$, and
			\emph{sort-decreasing}, i.e.,\
			for each  $t  \to  t'$ in
			$\overrightarrow{E_0}$ and substitution $\sigma$, {$t'\sigma \in \Tc_{\Sigma}(\Xc)_{\sort{s}}$ implies $t\sigma \in \Tc_{\Sigma}(\Xc)_{\sort{s}}$}
		\end{enumerate}
		\label{def:decomposition}
	\end{definition}

\noindent
In the following, we often abuse notation and say that
$(\Symbols,B,E_0)$ is a decomposition of an order-sorted equational theory
$\caE=(\Symbols,E)$ even if $E \neq E_0 \uplus B$ but 
$E_0$ is instead the explicitly extended $B$-coherent completion of a set $E_0'$ such that 
$E = E_0' \uplus B$.

\subsection{Pure homeomorphic embedding}

The pure  (syntactic) homeomorphic embedding relation known from term algebra \cite{Kru60}
was introduced by Dershowitz for 
variable-arity symbols in \cite{DJ90} and for fixed-arity symbols in \cite{Dershowitz1979}. In the following, we consider only fixed-arity symbols.

\begin{definition}[Homeomorphic embedding, Dershowitz \cite{Dershowitz1979}] \label{def:der79}   
	The  homeomorphic embedding relation $\blacktrianglelefteq$ over 
	$\Tc_{\Sigma}$
	is defined as follows:
	
	\begin{center}
		\begin{tabular}{C{5cm} C{5cm}}	
			{\large $\frac{\exists i \in\left\{ 1,\ldots,n\right\} ~:~s~\blacktrianglelefteq ~t_{i}} {s ~\blacktrianglelefteq ~f(t_{1},\dots,t_{n})}$} & {\large $\frac{\forall i \in\left\{ 1,\ldots,n\right\} ~:~s_{i}~\blacktrianglelefteq~ t_{i}}{f(s_{1},\dots,s_{n}) ~\blacktrianglelefteq ~f\left(t_{1},\dots,t_{n}\right)}$}\\
		\end{tabular}
	\end{center}
with $n \geq 0$.
\label{def:dershowitz}
\end{definition}

Roughly speaking, the left inference rule deletes subterms, 
while the right inference rule  deletes context. 
We write $s \blacktrianglelefteq t$ if $s$ is derivable from $t$ using the above rules.
When $s \blacktrianglelefteq t$, we say that $s$ is (syntactically) \textit{embedded} in $t$ (or $t$ syntactically \textit{embeds} $s$).
Note that $\equiv \;\subseteq\;  \blacktrianglelefteq$,
 where  
 $\equiv$ denotes syntactic identity.

A well-quasi ordering $\preceq$ is a transitive and reflexive
binary relation such that, for
any infinite sequence 
of terms $t_1, t_2,\ldots$ with a finite number of operators,
there exist $j,k$ with $j < k$ and $t_j 
\preceq t_k$. 
\begin{theorem}[Tree Theorem, Kruskal \cite{Kru60}] \label{def:kruskal}  
	The embedding relation $\blacktrianglelefteq$ is a well-quasi-ordering on 
	$\Tc_{\Sigma}$.
\end{theorem}

The derivability relation given by 
 $\blacktrianglelefteq$ 
is 
mechanized in \cite{Middeldorp1995} by 
introducing the following term rewriting system $Emb(\Sigma)$ as follows: $t \blacktrianglelefteq t'$ if and only if $t' \rightarrow^*_{Emb(\Sigma)} t$.

\begin{definition}[Homeomorphic embedding rewrite rules, Middeldorp \cite{Middeldorp1995}]\label{def:generated-middeldorp}
	Let $\Sigma$ be   a signature. The  homeomorphic embedding   can be decided by the TRS $Emb(\Sigma)$ that   consists of all rewrite rules $$f(X_1, \cdots, X_n) \rightarrow X_i$$
	where $f \in \Sigma$ is a function symbol of arity  $n \geq 1$ and $i \in \{1, \cdots, n\}$.
\end{definition}	

Definition~\ref{def:der79} can be applied to terms of $\Tc_{\Sigma}(\Xc)$ by simply regarding the variables in terms as constants. However, this definition cannot be used when existentially quantified variables are considered. 
The following definition from \cite{Leuschel98,SG95} adapts the pure (syntactic)  homeomorphic embedding from \cite{DJ90} by adding a simple treatment of logical  variables where all variables are treated as if they were identical, which is enough for many symbolic methods such as the   partial evaluation of \cite{ACEM17}.  
Some extensions of $\trianglelefteq$ dealing with  varyadic symbols and infinite signatures are investigated in	\cite{Leu02}.

\begin{definition}[Variable-extended homeomorphic embedding, Leuschel  \cite{Leuschel98}] \label{def:emb-orig}
	The extended homeomorphic embedding relation $\trianglelefteq$ over 
	$\Tc_{\Sigma}(\Xc)$
	is defined in Figure~\ref{fig:emb},
where the Variable inference rule 
allows dealing with free (unsorted) variables in terms,
while the Diving and Coupling inference rules are equal to the pure (syntactic) homeomorphic embedding definition.
\end{definition}

\begin{figure}[!h]
	\begin{center}
		\begin{tabular}{C{3cm} C{3cm} C{5cm}}
			Variable & Diving & Coupling\\[1ex]
			{\large  $\frac{}{x \, \trianglelefteq \, y}$} & {\large $\frac{\exists i \in\left\{ 1,\ldots,n\right\} ~:~s\trianglelefteq \, t_{i} }{s\trianglelefteq f\left(t_{1},\ldots,t_{n}\right)}$} & {\large $\frac{\forall i \in\left\{ 1,\ldots,n\right\} ~:~s_{i}\trianglelefteq \, t_{i} }{f\left(s_{1},\ldots,s_{n}\right)\trianglelefteq f\left(t_{1},\ldots,t_{n}\right)}$}\\
		\end{tabular}
	\end{center}
	\caption{Variable-extended homeomorphic embedding}
	\label{fig:emb}
\end{figure}	

The extended embedding relation $\trianglelefteq$ is a well-quasi-ordering on the set of terms
$\Tc_{\Sigma}(\Xc)$ \cite{Leuschel98,SG95}. 
An alternative characterization without the hassle of explicitly handling variables can be  proved as follows.

\begin{lemma}[Variable-less characterization of $\trianglelefteq$] \label{lem:emb-novar}
Given a signature $\Sigma$, let $\Sigma^\sharp$ be an extension of $\Sigma$ with a new constant $\sharp$, and  let
  $t^\sharp$ denote the (ground) instance of $t$ where all variables have been replaced by $\sharp$.
Given two terms $t_1$ and $t_2$,
$t_1 \trianglelefteq t_2$ iff $t_1^\sharp \trianglelefteq t_2^\sharp$ iff $t_1^\sharp \blacktrianglelefteq t_2^\sharp$.
\end{lemma}

Moreover, Lemma \ref{lem:emb-novar} above allows the variable-extended relation  $\trianglelefteq$ of Definition~\ref{def:emb-orig} to be mechanized in a way similar to the rewriting relation $\rightarrow^*_{Emb(\Sigma)}$   used in 
Definition~\ref{def:generated-middeldorp} for the embedding $\blacktrianglelefteq$  of Definition~\ref{def:der79}:
$t_1 \trianglelefteq t_2$ if and only if $t_2^\sharp \rightarrow^*_{Emb(\Sigma^\sharp)} t_1^\sharp$.
By abuse of notation, from now on, we will  indistinctly consider either terms with variables or ground terms with $\sharp$,   whenever one formulation is simpler than the other.


\section{Homeomorphic embedding modulo equational axioms}\label{sec:lopstr}

The following definition given in \cite{ACEM17}
extends  the ``syntactically simpler'' homeomorphic embedding relation  
on nonground terms to the order-sorted case {\em modulo} a set of axioms $B$. The (order-sorted) relation $\trianglelefteq_{B}$ is 
called $B$--embedding (or embedding modulo $B$).
We define $v \;{\stackrel{ren}{=}}_B v'$ iff there is a renaming substitution $\sigma$ for $v'$ 
such that $v =_{B} v'\sigma$. 

\begin{definition}[(Order-sorted) homeomorphic embedding modulo $B$]\label{def:embeddingB}
	We define the 
	\linebreak $B$--embedding relation $\trianglelefteq_{B}$  (or embedding modulo $B$) as    $({\stackrel{ren}{=}}_B) . (\trianglelefteq) . ({\stackrel{ren}{=}}_B)$.
\end{definition}

\begin{example} \label{ex:nat-order-sorted}
	Consider the following rewrite theory (written in Maude syntax) 
	that defines the signature of natural numbers, with sort {\tt Nat} and   constructor operators \verb|0|, and \verb|suc| for sort \verb|Nat|. We also define the associative and commutative addition operator symbol \verb|_+_|.	

	{\small	
		\begin{verbatim}
		fmod NAT is 
		  sort Nat .
		  op 0 : -> Nat .
		  op suc : Nat -> Nat .	
		  op _+_ : Nat Nat -> Nat [assoc comm] .
		endfm
		\end{verbatim}
	}

	Then, we have $+(1, X{:}Nat) \trianglelefteq_B +(Y{:}Nat,+(1,3))$ because $+(Y{:}Nat,+(1,3))$ is equal to $+(1,+(Y{:}Nat,3))$ modulo associativity and commutativity,
	and $+(1, X{:}Nat) \trianglelefteq +(1,+(Y{:}Nat,3))$.
\end{example}

	The following result extends Kruskal's Tree Theorem 
	for the equational theories considered in this paper. We have to restrict it to  the class of finite equational theories in order to prove the result. $\cal B$ is called \emph{class-finite} if all $\cal B$-equivalence classes are finite. This includes the class of permutative equational theories. An equational theory $\caE$ is permutative if for all terms $t, ~t'$, the fact that $t =_\caE t'$ implies that the terms $t$ and $t'$ contain the same symbols with the same number of occurrences \cite{BHS89}. Permutative theories include any theory with any combination of symbols obeying any combination of associativity and commutativity axioms. 

\begin{theorem} \label{kruskal}
	For class-finite theories,
	the embedding relation  $\trianglelefteq_{B}$ is a well-quasi ordering of the set 
	$\Tc_{\Sigma}(\Xc)$ 
	for finite $\Sigma$, that is, 
	$\trianglelefteq_{B}$ is a quasi-order.
\end{theorem}

Function symbols with variable arity  are sometimes seen as associative operators. Let us briefly discuss the homeomorphic embedding modulo axioms   $\trianglelefteq_B$ of Definition \ref{def:embeddingB} in comparison to
 the variadic extension $\blacktrianglelefteqvar$ of Definition \ref{def:der79} as given in \cite{DJ90}:

	\begin{center}
		\large	
		\begin{tabular}{C{3cm} C{8cm}}
			{\small Diving} & {\small Coupling}\\[1ex]	
			{$\frac{\exists i \in\left\{ 1,\ldots,n\right\} ~:~s~{\blacktrianglelefteqvar} ~t_{i} }{s ~\blacktrianglelefteqvar ~f(t_{1},\dots,t_{n})}$} & {$\frac{\forall i \in\left\{ 1,\ldots,m\right\} ~:~s_{i}~\blacktrianglelefteqvar~ t_{j_i} , \mbox{\small with } 1 \leq j_1 < j_2 < \cdots < j_m \leq n }{f(s_{1},\dots,s_{m}) ~\blacktrianglelefteqvar ~f\left(t_{1},\dots,t_{n}\right)}$}
		\end{tabular}
	\end{center}
		
\begin{example} 
	Consider a variadic version of the addition symbol $\pr{+}$   of Example~\ref{ex:nat-order-sorted} 
	that allows any number of natural numbers to be used as arguments; for instance, $+(1, 2,3)$. On the one hand, $+(1)  \blacktrianglelefteqvar +(1, 2,3)$
	whereas $+(1) \not \trianglelefteq_B +(1,2,3)$,  
	with $B$  consisting of the 
	associativity and commutativity axioms for the  operator $\pr{+}$  
	(actually, $+(1)$ is ill-formed).   
	On the other hand,  we have both $+(1,2)   \blacktrianglelefteqvar +(1,0,3,2)$ 
	and $+(1,2) \trianglelefteq_B +(1,0,3, 2)$. 
	This is because any well-formed term  that consists of the addition (in any order) of the constants $0$, $1$, $2$, and $3$ (for instance,\ $+(+(1,0),+(3,2)$)  
	can be given a flat representation  
	$+(1,0,2,3)$. Note that there are many other equivalent terms, e.g., $+(+(1,2),+(3,0))$ or $+(+(1,+(3,2)), 0)$, 
	all of  which are represented
	by  the flattened term $+(0,1,2,3)$. Actually, because of the  associativity and  commutativity of symbol  
	$\pr{+}$,
	flattened terms like $+(1,0,2,3)$ can be further simplified into a single\footnote{
		Maude uses a term lexicographic order for the arguments of flattened terms \cite{EkerJAR02}.} {\em canonical representative} $+(0,1,2,3)$, hence also $+(1,2) \trianglelefteq_B +(0,1,2,3)$.
	A more detailed explanation of flat terms can be found in Section \ref{sec:cinco}.
	However, note that 
	$+(2,1) \trianglelefteq_B +(1,0, 3, 2)$ but $+(2, 1) \not \blacktrianglelefteqvar +(1,0,3,2)$ because the $\blacktrianglelefteqvar$ does not  consider the commutativity of symbol $+$.
\end{example}

Roughly speaking, in the worst case, the  homeomorphic embedding modulo axioms  of 
Definition~\ref{def:embeddingB}, $t \trianglelefteq_B t'$,  amounts to considering
all the elements in the $B$-equivalence classes of $t$ and $t'$ and then  checking for standard homeomorphic embedding,  $u \trianglelefteq u'$, every  pair $u$ and $u'$ of such  terms, one term from each class.
According to Definition~\ref{def:generated-middeldorp}, checking $u \trianglelefteq u'$ 
essentially boils down to the reachability analysis given by  $u' \rightarrow^*_{Emb(\Sigma)} u$.
Unfortunately, the enumeration of all terms in a $B$-equivalence class is impractical, as shown in the following example.

\begin{example}\label{ex:assoc-comm-pre}
Consider the  AC binary symbol $+$ of Example~\ref{ex:nat-order-sorted}  and the terms $t=+(1, 2)$ and $t'=+(2,+(3,1))$. 
The AC-equivalence class of $t$ contains two terms whereas the AC-equivalence class of $t'$ contains nine terms. This implies computing eighteen  reachability problems $u' \rightarrow^*_{Emb(\Sigma)} u$ in order to decide $t \trianglelefteq_{AC} t'$, in the worst case. Moreover, we   know a priori that half of these reachability tests 
will fail (those in which $1$ and $2$ occur in different   order in $u'$ and $u$;
for instance  $u'=+(1,+(2,3))$ and $u=+(2,1)$.
\end{example}

A more effective rewriting characterization of $\trianglelefteq_{B}$ can be achieved  by  lifting Definition \ref{def:generated-middeldorp} to 
the order-sorted   and   \emph{modulo} case in a natural way.
However, 
ill-formed terms can be produced by na\"{\i}vely applying the rules  
$f(X_1,\ldots,X_n ) \rightarrow X_i$ of Definition \ref{def:generated-middeldorp}
 to typed (i.e.,\ order-sorted) terms. For example, $``{\tt (0 \leq 1) ~or ~true" \rightarrow ``0}$
${\tt or ~true}"$.

In the order-sorted  context  we can   overcome this drawback   as follows. Assume that $\Sigma$ has no ad-hoc overloading.
Then, we can   extend $\Sigma$ to a new signature $\Sigma^{\cal U}$ by  adding a new top sort $\cal U$ that is bigger than all other sorts.
Now, for each $f: A_{1}, \ldots, A_{n} \rightarrow A$ in $\Sigma$, we add the rules $f(X_{1}{:}{\cal U}, \ldots, X_{n}{:}{\cal U}) \rightarrow X_i {:}{\cal U}$, $1\leq i\leq n$.
In this way, rewriting with $\rightarrow^{*}_ {
	{Emb}(\Sigma^{\cal U})
	/B}$ becomes a relation between well-formed $\Sigma^{\cal U}$-terms, as first proposed in \cite{ACEM17}.

\begin{definition}[(Order-sorted) homeomorphic embedding rewrite rules modulo $B$ \cite{ACEM17}]\label{def:lopstr}
	Let 
	$(\Sigma,B,\overrightarrow{E_0})$ be an equational theory decomposition. Let us introduce the following signature transformation {$\Sigma \ni (f : \sort{s}_{1} \dots \sort{s}_{n} \rightarrow \sort{s}) \mapsto
	(f : {\cal U}  \stackrel{n}{...} {\cal U} \rightarrow  {\cal U}) \in \Sigma^{u}$}, where {$\cal U$} conceptually represents a universal supersort of all sorts in $\Sigma$. Also, for any $\Sigma$-term $t$,  $t^{u}$ leaves the term $t$  unchanged
	but regards all its variable  as unsorted (i.e., of sort  {$\cal U$}). We define the TRS  {Emb}($\Sigma$) that consists of all rewrite rules.
	$$
	f(X_1 {:} {\cal U},\ldots,X_n {:} {\cal U}) \rightarrow X_i {:} {\cal U}
	$$
	for each $f : A_1, \ldots, A_n \rightarrow A$  in $\Sigma$ and $i\in\{1,\ldots,n\}$.	
\end{definition}

In the sequel, we consider equational theories $B$  that may contain any combination of associativity and/or commutativity axioms for any binary symbol in the signature.
Also, for the sake of simplicity we often omit 
sorts 
when no confusion can arise.
\begin{proposition}
	Given $\Sigma$ and  $B$,  for $t$ and $t'$ in $\Tc_{\Sigma}(\Xc)$, $t ~\trianglelefteq_{B} ~t'$
	~iff~ $(t'^u)^\sharp\rightarrow^{*}_ {
		{Emb}((\Sigma^{\cal U})^\sharp)
		/B} (t^u)^\sharp$.
	\label{prop:embVar}
\end{proposition}
	
\begin{example} \label{ex:nat-top}
	Consider the order-sorted signature for natural numbers 
	of  Example \ref{ex:nat-order-sorted}. Let us represent by sort \pr{U}  in Maude  the unique (top) sort  of the transformed signature: 
	
	{\small
		\begin{verbatim}
		fmod NAT-U is 
		  sort U .
		  op 0 : -> U .
		  op suc : U -> U .	
		  op _+_ : U U -> U [assoc comm] .
		endfm
		\end{verbatim}
	}

	Likewise, the terms expressed in $\Sigma$  must also be   transformed to be expressed as $\Sigma^{\cal U}$-terms. For instance, given the $\Sigma$-terms $t= \texttt{X:Nat}$\footnote{The expression $X{:}S$ represents an explicit definition of a variable $X$ of sort $S$ in Maude.} and $t'=\texttt{suc(Y:Nat)}$, the corresponding     $\Sigma^{\cal U}$-terms are $t=\texttt{X:U}$ and $\texttt{suc(Y:U)}$, respectively. 
	
	The associated TRS ${Emb}(\Sigma)$ contains the following two rules for the operator \pr{+}:
			\begin{align}
			+(X_1 {:}{U},X_2{:}{U}) \rightarrow X_1 {:}{U}\nonumber\\
			+(X_1 {:} {U},X_2 {:} {U}) \rightarrow X_2 {:} {U}\nonumber
			\end{align}		
	
	However, since    the rules of ${Emb}(\Sigma)$ are applied modulo the commutativity   of symbol  {\pr{+}}, 
	in practice, we can get rid of either of the two  rules above since only one is required
	in Maude.
\end{example}

\begin{example}\label{ex:assoc-comm-pre2}
	Following Example~\ref{ex:assoc-comm-pre},
	instead of comparing pairwisely all  terms in the equivalence classes of $t$ and $t'$, we
	choose  ${Emb}(\Sigma)$ to contain just the rewrite rule   
	 $+(X_1 {:} {U},X_2 {:} {U}) \rightarrow X_2 {:} {U}$, we use it to prove the rewrite step 
	$+(2,+(3,1)) \rightarrow_{{Emb}(\Sigma)/B}  
	  +(2,1)$, and finally we check that $+(2,1) =_B +(1,2)$, with $B=\{A,C\}$.
	However, there are six alternative rewriting steps stemming from the initial term $+(2,+(3,1))$, all of which result from applying the very 
	same rewrite rule above to the term (modulo AC), five of which are useless for proving the considered  embedding (the selected redex is underlined):
	
	$$
	{\small
	\begin{array}{lll} 
	+(2,\underline{+(3,1)})\rightarrow_{{Emb}(\Sigma)/B} +(2,1)\ \ \  \nonumber
	&
	+(2,\underline{+(3,1)})\rightarrow_{{Emb}(\Sigma)/B} +(2,3)\ \ \ \nonumber
	&
	\underline{+(2,+(3,1)}) \rightarrow_{{Emb}(\Sigma)/B} +(3,1)\nonumber\\[0.7ex]
	\underline{+(2,+(3,1)}) \rightarrow_{{Emb}(\Sigma)/B} 1~~~~~~~~~~\nonumber
	&
	\underline{+(2,+(3,1)}) \rightarrow_{{Emb}(\Sigma)/B} 2~~~~~~~~~~\nonumber
	&
	\underline{+(2,+(3,1)}) \rightarrow_{{Emb}(\Sigma)/B} 3~~~~~~~~~~\nonumber
	\end{array}
	}$$
	
	\noindent
	For a term with $k$ addends, we have $(2^k)-2$ rewriting steps. This leads to a huge combinatorial explosion when considering the complete rewrite search tree.
\end{example}

Moreover, there are three problems with Definition \ref{def:lopstr}. 
First, the intrinsic non-determinism of the rules may unnecessarily produce  an extremely  large  search space. 
Second,   as shown in Example \ref{ex:assoc-comm-pre2}, this intrinsic non-determinism in the presence of axioms is intolerable, that is, unfeasible to handle.
Third, the associated
reachability problems do not scale up to complex embedding problems so that  a suitable search strategy  must be introduced.  
We address these problems stepwisely in the sequel.


\section{Goal-driven homeomorphic embedding modulo $B$}\label{sec:gd}
The formulation of homeomorphic embedding as a reachability problem 
by using the rewrite rules of Definition~\ref{def:lopstr} generates a blind search that  does not take advantage of the   actual terms $t$ and $t'$  
being compared for embedding. In this section, we provide a  more refined formulation of homeomorphic embedding modulo axioms that is \emph{goal driven}
in the sense that, given an  embedding problem (or \emph{goal}), $t ~\trianglelefteq_{B} ~t'$,  it inductively processes the terms  $t$ and $t'$ 
in a top-down manner. 

First, we introduce in the following section a  
calculus 	that extends the  
 homeomorphic embedding  relation of Definition \ref{def:emb-orig} to the order-sorted equational case.

\subsection{An homeomorphic embedding calculus modulo $B$}
Let us introduce a calculus for embeddability goals  $t \trianglelefteqgd_B t'$
that directly handles  in the deduction system  the algebraic axioms of $B$, with $B$ being any combination of A and/or C axioms for the theory operators.
Roughly speaking, this is achieved by specializing w.r.t.\ $B$ the coupling rule  of Definition \ref{def:emb-orig}.  

\begin{definition}[Goal-driven homeomorphic embedding modulo $B$] \label{def:emb-acu}
The homeomorphic embedding relation modulo $B$ is defined as the smallest relation that satisfies the  inference rules of Definition \ref{def:emb-orig} together with the new inference rules given in Figure \ref{fig:gd}. That is:

\begin{enumerate}	
	\item the three inference rules (Variable, Diving, and Coupling) of Definition~\ref{def:emb-orig}
		for any function symbol;
	\item one
	extra coupling rule for the case of a commutative symbol with or without associativity (Coupling$_{C}$);
	\item  
	two extra coupling rules for the case of an associative symbol with or without commutativity (Coupling$_{A}$);
	and 
	\item
	two extra coupling rules for the case of an associative-commutative symbol (Coupling$_{AC}$).
\end{enumerate}

\begin{figure}[h!]
	{\small
		\[
			\textbf{Coupling$_C$}~~~~
		\frac
		{
			s_{0}\trianglelefteqgd_{B}t_{1}
			\ \ \wedge\ \ 
			s_{1}\trianglelefteqgd_{B}t_{0}
		}
		{
			f(s_{0},s_{1})\trianglelefteqgd_{B} f(t_{0},t_{1})	
		}
		\]
	}

	{\small
		\[
			\textbf{Coupling$_A$}~~~~
		\frac
		{
			f(s_{0},s_{1})\trianglelefteqgd_{B}t_{0}
			\ \ \wedge\ \ 
			s_{2}\trianglelefteqgd_{B}t_{1}
		}
		{
			f(s_{0},f(s_{1},s_{2}))\trianglelefteqgd_{B} f(t_{0},t_{1})			
		}
		\hspace{5ex}
		\frac
		{
			s_{0}\trianglelefteqgd_{B}f(t_{0},t_{1})
			\ \ \wedge\ \ 
			s_{1}\trianglelefteqgd_{B}t_{2}
		}
		{
			f(s_{0},s_{1})\trianglelefteqgd_{B} f(t_{0},f(t_{1},t_{2}))
		}
		\]
	}

	{\small
		\[
			\textbf{Coupling$_{AC}$} ~~~~
		\frac
		{
			f(s_{0},s_{1})\trianglelefteqgd_{B}t_{1}
			\ \ \wedge\ \ 
			s_{2}\trianglelefteqgd_{B}t_{0}
		}
		{
			f(s_{0},f(s_{1},s_{2}))\trianglelefteqgd_{B} f(t_{0},t_{1})
		}
		\hspace{4ex}
		\frac
		{
			s_{1}\trianglelefteqgd_{B}f(t_{0},t_{1})
			\ \ \wedge\ \ 
			s_{0}\trianglelefteqgd_{B}t_{2}
		}
		{
			f(s_{0},s_{1})\trianglelefteqgd_{B} f(t_{0},f(t_{1},t_{2}))
		}
		\]
	}
	\caption{Extra coupling rules for A, C, AC symbols}
	\label{fig:gd}
\end{figure}

\end{definition}

\begin{proposition}
	Given $\Sigma$ and  $B$,  for terms $t$ and $t'$ in $\Tc_{\Sigma}(\Xc)$, 
	$t \trianglelefteq_B t'$ iff $t \trianglelefteqgd_B t'$ . 
	\label{prop:embGoalDriven}
\end{proposition}

\begin{example}\label{ex:assoc-comm}
	Consider the binary symbol $+$ 
	obeying associativity and commutativity axioms,  and the terms $t=+(1, 2)$ and $t'=+(2,+(3,1))$ of
	Example~\ref{ex:assoc-comm-pre2}. 
		We can prove  $t \trianglelefteqgd_B t'$ by 
	
	\[
	\frac
	{
	\frac{1 \trianglelefteqgd_B 1}
	{1 \trianglelefteqgd_B +(3,1)} \hspace{4ex} 2 \trianglelefteqgd_B 2}
	{+(1,2) \trianglelefteqgd_B +(2,+(3,1))}
	\]	
	
	We can also prove a more complex embedding goal  by first using the right inference rule for AC of Figure~\ref{fig:gd} and then  the generic Coupling and Diving inference rules.
	\[
	\frac
	{
	\frac
	{
	\frac
	{2 \trianglelefteqgd_B 2}
	{2 \trianglelefteqgd_B +(4,2)}
	\hspace{2ex} 3 \trianglelefteqgd_B 3 } 
	{+(2,3) \trianglelefteqgd_B +(+(4,2),3)} 
	\hspace{4ex} 
	1 \trianglelefteqgd_B 1}
	{+(1,+(2,3)) \trianglelefteqgd_B +(+(4,2),+(3,1))}
	\]	
	It is immediate to see that, when the size of the involved terms  $t$ and $t'$ grows,  the improvement in performance of $\trianglelefteqgd_B$ w.r.t. $\trianglelefteq_B$ can be significant  (just compare these two embedding proofs with the corresponding search trees for  $\trianglelefteq_B$).
\end{example}

\subsection{Reachability-based, goal-driven homeomorphic embedding formulation}

Let us provide  a more operational goal-driven 
characterization of the   homeomorphic embedding modulo $B$. We formalize it in the reachability style of Definition~\ref{def:lopstr}.
The main challenge  here  is    how to generate a suitable rewrite theory $R^{rogd}(\Sigma,B)$ that can decide 
embedding modulo 
$B$  
by running a reachability goal.

\begin{definition}[Goal-driven homeomorphic embedding rewrite rules modulo $B$]\label{gd-theory}
	Given $\Sigma$ and  $B$,
	we	define the TRS $R^{rogd}(\Sigma,B)$ as follows.
	\begin{enumerate}
		\item We include in $R^{rogd}(\Sigma,B)$ a rewrite rule of the form 
		$u \trianglelefteqrogd_{B} v \to true$ for each
		 (particular intance of the)   inference rules of the form 
		$\frac{}{u \trianglelefteqgd_{B} v}$ given
		  Definition~\ref{def:emb-acu}
		(e.g., the Variable Inference Rule from Definition~\ref{def:emb-orig}
		or  the Coupling Inference Rule from Definition~\ref{def:emb-orig}, for the case of a constant symbol $c$).
		
		\item We include in $R^{rogd}(\Sigma,B)$ a rewrite rule of the form 
		$u \trianglelefteqrogd_{B} v \to u_1 \trianglelefteqrogd_{B}v_1 \wedge \cdots \wedge u_k \trianglelefteqrogd_{B}v_k$
		for each 
		 (particular intance of the) inference rules 
		of the form 
		$\frac{u_1 \trianglelefteqgd_{B}v_1 \wedge \cdots \wedge u_k \trianglelefteqgd_{B}v_k}{u \trianglelefteqgd_{B} v}$
		given in Definition~\ref{def:emb-acu}.
	\end{enumerate}
\end{definition}	

\begin{proposition}\label{prop:3}
	Given $\Sigma$ and  $B$,  for terms $t$ and $t'$ in $\Tc_{\Sigma}(\Xc)$,  $t ~\trianglelefteqgd_{B} ~t'$
	~iff~ $(t \trianglelefteqrogd_{B}  t')\rightarrow^{\ast}_{{R^{rogd}(\Sigma,B)}/B} true$. 
\end{proposition}

\begin{example}\label{ex:assoc-comm-nd}
		Consider 
		the binary symbol $+$ of Example~\ref{ex:nat-order-sorted}.
		According to Definition~\ref{def:emb-acu}, there are twelve  inference rules for  $\trianglelefteqgd_{B}$:
	\begin{center}
		\begin{tabular}{C{3cm} C{3cm} C{3cm}}	
			\large	
			{\small Variable} & {\small Diving} & {\small Coupling} \\[1ex]
			{\large  $\frac{}{x\trianglelefteqgd_{B} y}$} & {\large$\frac{x\, \trianglelefteqgd_{B} \, t_{1}} {x\, \trianglelefteqgd_{B} suc\left(t_{1}\right)}$} & {\large $\frac{}{0 \, \trianglelefteqgd_{B} 0}$}
			\\[2ex]	
			& {\large$\frac{x\, \trianglelefteqgd_{B} \, t_{1}} {x\, \trianglelefteqgd_{B} +\left(t_{1}, t_{2}\right)}$} & {\large $\frac{t_{1}\, \trianglelefteqgd_{B} \, t'_{1} }{suc\left(t_{1}\right) \, \trianglelefteqgd_{B}suc\left(t'_{1}\right)}$}
	\\[2ex]		
			& {\large$\frac{x\, \trianglelefteqgd_{B} \, t_{2}} {x\, \trianglelefteqgd_{B} +\left(t_{1}, t_{2}\right)}$} 				
			& {\large $\frac{t_{1}\, \trianglelefteqgd_{B} \, t'_{1} \ \wedge \ t_{2} \, \trianglelefteqgd_{B} \, t'_{2} }{+\left(t_{1},t_{2}\right) \, \trianglelefteqgd_{B}+\left(t'_{1},t'_{2}\right)}$}
			\\[2ex]
		\end{tabular}
	\end{center}
	
		\begin{center}
		\begin{tabular}{C{3cm} C{4cm} C{4cm}}	
			\large	
			{\small Coupling$_C$} & {\small Coupling$_A$} & {\small Coupling$_{AC}$}\\[1ex]
			 {\large $\frac{t_{1}\, \trianglelefteqgd_{B} \, t'_{2} \ \wedge \ t_{2} \, \trianglelefteqgd_{B} \, t'_{1} }{+\left(t_{1},t_{2}\right) \, \trianglelefteqgd_{B}+\left(t'_{1},t'_{2}\right)}$}
			& {\large $\frac{+(t_{0},t_{1})\, \trianglelefteqgd_{B} \, t'_{1} \ \wedge \ t_{2} \, \trianglelefteqgd_{B} \, t'_{2} }{+\left(t_{0},+(t_{1},t_{2})\right) \, \trianglelefteqgd_{B}+\left(t'_{1},t'_{2}\right)}$}
			& {\large $\frac{+(t_{0},t_{1})\, \trianglelefteqgd_{B} \, t'_{2} \ \wedge \ t_{2} \, \trianglelefteqgd_{B} \, t'_{1} }{+\left(t_{0},+(t_{1},t_{2})\right) \, \trianglelefteqgd_{B}+\left(t'_{1},t'_{2}\right)}$}
			\\[2ex]	
			& {\large $\frac{t_{1}\, \trianglelefteqgd_{B} \, +(t'_{0},t'_{1}) \ \wedge \ t_{2} \, \trianglelefteqgd_{B} \, t'_{2} }{+\left(t_{1},t_{2}\right) \, \trianglelefteqgd_{B}+(t'_{0},+(t'_{1},t'_{2}))}$}
			& {\large $\frac{t_{2}\, \trianglelefteqgd_{B} \, +(t'_{0},t'_{1}) \ \wedge \ t_{1} \, \trianglelefteqgd_{B} \, t'_{2} }{+\left(t_{1},t_{2}\right) \, \trianglelefteqgd_{B}+(t'_{0},+(t'_{1},t'_{2}))}$}
	\\[2ex]		
		\end{tabular}
	\end{center}
	
	However, the corresponding TRS $R^{rogd}(\Sigma,B)$ only contains six rewrite rules because, due to pattern matching modulo associativity and commutativity in rewriting logic, 
	the other rules are redundant: 
	
	$$
	\begin{array}{rrll}
	\mbox{\small (Diving)}  & x &\trianglelefteqrogd_{B} suc(T_{1})&\rightarrow 	x \trianglelefteqrogd_{B}T_1 \hspace{5ex} \\
	&x &\trianglelefteqrogd_{B}+\left(T_{1},T_{2}\right) &\rightarrow 	x   \trianglelefteqrogd_{B}T_1 
	\\
	\mbox{\small (Coupling)} &\sharp &\trianglelefteqrogd_{B}\sharp&\rightarrow true	   \ \\
	&0 &\trianglelefteqrogd_{B}0&\rightarrow true	   \ \\
	&suc(T_1) &\trianglelefteqrogd_{B}suc\left(T'_{1}\right) &\rightarrow 	  T_1\trianglelefteqrogd_{B}T'_1 \ \\
	\mbox{\small (Coupling$_{\emptyset,C,A,AC}$)} &+(T_1,T_{2}) &\trianglelefteqrogd_{B}+\left(T'_{1},T'_{2}\right) &\rightarrow 	  T_1\trianglelefteqrogd_{B}T'_1 \ \wedge\  T_2 \trianglelefteqrogd_{B} T'_{2}\\
	\end{array}
	$$
	
	\noindent
	For example, the rewrite sequence proving $+(1,+(2,3)) \trianglelefteqrogd_B +(+(4,2),+(3,1))$ is: 
	
	\begin{align}
	 +(1,+(2,3)) \trianglelefteqrogd_B +(+(4,2),+(3,1))
	& \to_ {{R^{rogd}(\Sigma,B)}/B}
	+(2,3)) \trianglelefteqrogd_B +(+(4,2),3)
	\wedge
	1 \trianglelefteqrogd_B 1\nonumber\\[-1ex]
	&\to_ {{R^{rogd}(\Sigma,B)}/B}
	2 \trianglelefteqrogd_B +(4,2)
	\wedge
	3 \trianglelefteqrogd_B 3
	\nonumber\\[-1ex]
	&\to_ {{R^{rogd}(\Sigma,B)}/B}
	2 \trianglelefteqrogd_B 2
	\nonumber\\[-1ex]
	&\to_ {{R^{rogd}(\Sigma,B)}/B} true
	\nonumber
	\end{align}
	
	\noindent
	Although the improvement in performance 
	achieved by using the rewriting relation
	$\to_ {{R^{rogd}(\Sigma,B)}/B}$
	versus
	the rewriting relation
	$\rightarrow^{*}_ {
		{Emb}(\Sigma)/B
		}$
	is important, the search space is still huge
	since 
	the expression 
	$+(1,+(2,3)) \trianglelefteqgd_B +(+(4,2),+(3,1))$ matches the left-hand side $+(T_1,T_2) \trianglelefteqgd_B +(T'_1,T'_2)$ in many different ways
	(e.g.,\
	$\{T_1 \mapsto 1, T_2 \mapsto +(2,3), \ldots \}$,
	$\{T_1 \mapsto 2, T_2 \mapsto +(1,3), \ldots \}$,
	$\{T_1 \mapsto 3, T_2 \mapsto +(1,2), \ldots \}$
	).
\end{example}

In the following section, we further optimize the calculus of   homeomorphic embedding   modulo axioms by 
 considering equational (deterministic) normalization (thus avoiding search) 
and by exploiting the meta-level features of Maude (thus avoiding any theory generation).


\section{Meta-Level  deterministic goal-driven homeomorphic embedding modulo $B$}
\label{sec:cinco}

The meta-level representation
of terms in Maude \cite[Chapter 14]{maude-book} works with  flattened versions of the terms
that are rooted by poly-variadic versions of  the 
associative (or associative-commutative) symbols. For instance,
given an  associative (or associative-commutative) symbol  $f$ 
with $n$ arguments and $n\geq 2$,
flattened terms rooted by $f$ are canonical forms w.r.t. the set of rules given by the following rule schema
$$f(x_{1},\ldots,f(t_{1},\ldots,t_{n}),\ldots,x_{m})
\to f(x_{1},\ldots,t_{1},\ldots,t_{n},\ldots,x_{m}) \ \ \ n,m \geq 2$$
Given an  associative (or associative-commutative) symbol $f$ and a term
$f(t_{1},\ldots,t_{n})$,
we call \emph{$f$-alien terms} (or simply \emph{alien terms})
those terms among the $t_{1},\ldots,t_{n}$
that are not rooted by $f$.  
In the following, we implicitly consider that all terms are in $B$-canonical form. 

In the sequel, 
a variable $x$ of sort $\sort{s}$ is meta-represented as $\bar{x} = {\texttt{'}x}{:}\sort{s}$ and a non-variable term $t=f(t_1,\ldots,t_n)$, 
with $n \geq 0$, is meta-represented as  
$\bar{t} = {\texttt{'}\!f}[\bar{t_1}, \ldots, \bar{t_n}]$.

\begin{definition}[Meta-level homeomorphic embedding modulo $B$] \label{def:reflection-bool-emb-acu}
	The  meta-level homeomorphic embedding   modulo $B$,  
	  $\trianglelefteqdeml_B$,   is defined for term meta-representations by means of the equational theory $E^{ml}$ given in Figure~\ref{fig:meta-level}, 
	  where the auxiliary meta-level functions $\textbf{any}$ and $\textbf{all}$ implement the existential and universal tests in the Diving and Coupling inference rules of  Figure \ref{fig:emb}, 
	  and we introduce two new  meta-level  functions  $\textbf{all\_A}$ and $\textbf{all\_AC}$ that implement  existential tests that are specific to A and AC symbols. 
	  For the sake of readability, these new existential tests are also formulated (for ordinary terms instead of meta-level terms) as the inference rules Coupling$_A$ and 
	  Coupling$_{AC}$ of  Figure \ref{fig:ml}.
\end{definition}


{\linespread{0.9} 
	\begin{figure}[h!]
		$
		\begin{array}{rll}
		\sharp &\trianglelefteqdeml_{B} \sharp &= \textbf{true}\\
		F[TermList] &\trianglelefteqdeml_{B} \sharp &= \textbf{false}\\	
		T &\trianglelefteqdeml_{B} F[TermList] &= \textbf{any}(T,TermList) \hspace{16.7ex} \mbox{ if } root(T) \neq F\\
		F[TermList1] &\trianglelefteqdeml_{B} F[TermList2] &= \textbf{any}(F[TermList1],TermList2)\\
		&&\hspace{2ex} \textbf{or} 
		\  \textbf{all}(TermList1,TermList2)\\
		F[U, V] &\trianglelefteqdeml_{B} F[X, Y] &= \textbf{any}(F[U, V],[X, Y])  \hspace{5ex} \mbox{ ~~~~~~~~~~~~~~~~~if $F$ is $C$} \\
		&&\hspace{2ex} \textbf{or} 	
		(\ U\trianglelefteqdeml_{B}X \ \textbf{and}\  V\trianglelefteqdeml_{B}Y\ )\\
		&&\hspace{2ex} \textbf{or} 
		\ (\ U\trianglelefteqdeml_{B}Y \ \textbf{and}\  V\trianglelefteqdeml_{B}X\ )\\
		F[TermList1] &\trianglelefteqdeml_{B} F[TermList2] &= \textbf{any}(F[TermList1],TermList2) \hspace{4.9ex} \mbox{ if $F$ is $A$} \\
		&&\hspace{2ex} \textbf{or} 
		\  \textbf{all\_{A}}(TermList1,TermList2)\\
		F[TermList1] &\trianglelefteqdeml_{B} F[TermList2] &= \textbf{any}(F[TermList1],TermList2) \hspace{4.9ex} \mbox{ if $F$ is $AC$} \\
		&&\hspace{2ex} \textbf{or} 
		\ \textbf{all\_{AC}}(TermList1,TermList2)\\
		\end{array}\\ [2ex]
		\begin{array}{rl}
		\textbf{any}(U,nil) &= \textbf{false} \\
		\textbf{any}(U,V : L) &= U\trianglelefteqdeml_{B}V\ \textbf{or}\ \textbf{any}(U,L) \\[2ex]
		\textbf{all}(nil,nil) &= \textbf{true} \\
		\textbf{all}(nil,U : L) &= \textbf{false} \\
		\textbf{all}(U : L,nil) &= \textbf{false} \\
		\textbf{all}(U : L1,V : L2) &= U\trianglelefteqdeml_{B}V\ \textbf{and}\ \textbf{all}(L1,L2) \\[2ex]
		\textbf{all\_{A}}(nil,L) &= \textbf{true} \\
		\textbf{all\_{A}}(U : L,nil) &= \textbf{false} \\
		\textbf{all\_{A}}(U : L1,V : L2) &= (U\trianglelefteqdeml_{B}V\ \textbf{and}\ \textbf{all\_{A}}(L1,L2))\ \textbf{or}\ \textbf{all\_{A}}(U : L1,L2)) \\[2ex]
		\textbf{all\_{AC}}(nil,L) &= \textbf{true} \\
		\textbf{all\_{AC}}(U : L1,L2) &= \textbf{all\_{AC}\_Aux}(U : L1,L2,L2) \\
		\textbf{all\_{AC}\_Aux}(U : L1,nil,L3) &= \textbf{false} \\
		\textbf{all\_{AC}\_Aux}(U : L1,V : L2,L3) &= (U\trianglelefteqdeml_{B}V\ \textbf{and}\ \textbf{all\_{AC}}(L1,\textbf{remove}(V,L3)))\ \\ & ~~~~~ \textbf{or}\  \textbf{all\_{AC}\_Aux}(U : L1,L2,L3)) \\[2ex]
		\textbf{remove}(U,nil) &= \textbf{nil} \\
		\textbf{remove}(U,V : L) &= \textbf{if}\ U = V\ \textbf{then}\ L\ \textbf{else}\ V : \textbf{remove}(U,L)\\
		\end{array}
		$
		\caption{Meta-level  homeomorphic embedding modulo axioms}	
		\label{fig:meta-level}
	\end{figure}
}

\begin{example}
	Given the embedding problem for terms
	$+(1,+(2,3))$ and $+(+(4,2),+(3,1))$,
	the corresponding call to the meta-level homeomorphic embedding  $\trianglelefteqdeml_B$ of Definition \ref{def:reflection-bool-emb-acu} is 
	$\texttt{'}\!{+}[\texttt{'}1,\texttt{'}2,\texttt{'}3] \trianglelefteqdeml_B \texttt{'}\!{+}[\texttt{'}4,\texttt{'}2,\texttt{'}3,\texttt{'}1]$. 
\end{example}


\begin{figure}[h!]
	\[
		\textbf{Coupling$_A$}
	\frac{
		\exists j\in\left\{ 1,\dots,m-n+1\right\} : s_{1}\trianglelefteqdeml_{B}t_{j} \wedge f\left(s_{2},\ldots,s_{n}\right)\trianglelefteqdeml_{B}f\left(t_{j+1},\ldots,t_{m}\right) 
		\wedge \forall k<j: {s_{1}\not\trianglelefteqdeml_{B}t_{k}}
	}
	{
		f\left(s_{1},\ldots,s_{n}\right)\trianglelefteqdeml_{B}f\left(t_{1},\ldots,t_{m}\right) 
	}\]
	{
		\[
			\textbf{Coupling$_{AC}$}
		\frac{
			\exists j\in\left\{ 1,\dots,m\right\} : s_{1}\trianglelefteqdeml_{B}t_{j} \wedge f\left(s_{2},\ldots,s_{n}\right)\trianglelefteqdeml_{B}f\left(t_1,\ldots,t_{j-1},t_{j+1},\ldots,t_{m}\right)  
		}
		{
			f\left(s_{1},\ldots,s_{n}\right)\trianglelefteqdeml_{B}f\left(t_{1},\ldots,t_{m}\right)
		}
		\]
	}
	\caption{Coupling rule for associativity-commutativity functions}
	\label{fig:ml}
\end{figure}

\begin{proposition}\label{prop:metaLevel}
	Given $\Sigma$ and  $B$,  for terms $t$ and $t'$ in $\Tc_{\Sigma}(\Xc)$,
	$t \trianglelefteqgd_B t'$ iff $(t \trianglelefteqdeml_B t')!_{E^{ml}/B}=\textit{true}$. 
\end{proposition}

Finally, a further optimized version of Definition \ref{def:reflection-bool-emb-acu} can be easily defined by replacing the Boolean conjunction ({\em and}) and disjunction  ({\em or}) operators
with the computationally more efficient  Maude Boolean operators {\tt and-then}  and {\tt or-else} that avoid evaluating the second  argument when the result of evaluating the first one 
suffices to compute the result.

\begin{definition}[Strategic meta-level deterministic embedding modulo $B$] \label{def:bool-emb-acu}
We define  $\trianglelefteqdestrat_B$ as the strategic version of relation $\trianglelefteqdeml_B$ that is obtained by replacing the Boolean operators {\tt and}    and {\tt or} with Maude's {\tt and-then} 
operator for \emph{short-circuit} version  of conjunction and the {\tt or-else} operator for short-circuit disjunction \cite[Chapter 9.1]{maude-book}, respectively.
\end{definition}


\section{Experiments}

We have implemented in Maude all four equational homeomorphic   embedding formulations
$\trianglelefteq_B$, $\trianglelefteqrogd_B$, $\trianglelefteqdeml_B$, and $\trianglelefteqdestrat_B$
 of previous sections.
The implementation  consists of  approximately 250 function definitions (2.2K lines of Maude source code) and
 is publicly available online at 
\url{http://safe-tools.dsic.upv.es/victoria/jsp-pages/embedding.jsp}. 
In this section, we provide an experimental comparison of the four
equational homeomorphic embedding  implementations by running
a significant number of equational embedding goals. 
In order to compare the performance of the different implementations in the worst  possible scenario, all benchmarked  goals return false, which ensures that the 
whole search space for each goal has been completely explored,
while the execution times for succeeding goals whimsically depend on the  particular node of the search tree where  success is found.

We tested our implementations
on a 3.3GHz Intel Xeon E5-1660 with 64 GB of RAM running   Maude v2.7.1,  and we considered the average of ten executions for each test. 
	We have chosen four representative programs:
(i) 
{\it KMP}, the classical KMP string pattern matcher \cite{AFV98};
(ii)
{\it NatList}, a Maude implementation of lists of natural numbers;
(iii)
{\it Maze}, a non-deterministic Maude specification that defines a maze game in which multiple players must reach a given exit point by walking or jumping, 
where colliding players are eliminated from the game \cite{ABFS-JSC15};
and
(iv)
{\it Dekker}, a Maude specification that models a faulty version of Dekker's protocol, 
one of the earliest solutions to the mutual exclusion problem that appeared in \cite{maude-book}.
As testing benchmarks we considered a set of representative   embeddability problems  for  the four programs that are generated  during the execution of the partial evaluator Victoria \cite{ACEM17}.

Tables \ref{tbl:sizeTheory}, \ref{tbl:performance}, and \ref{tbl:natComparison} below analyze different aspects of the implementation.
In Table \ref{tbl:sizeTheory}, we compare the size  of the generated rewrite theories for the na\"{\i}ve and the goal-driven definitions  versus the meta-level definitions. For both, $\trianglelefteqdeml_B$ and $\trianglelefteqdestrat_B$,  there are the same number  (21) of generated equations ($\sharp$\pr{E}),
whereas   the number  of generated  rules ($\sharp$\pr{R}) is zero  because both definitions are purely equational (deterministic) and just differ in the version of the boolean 
operators being used. As for the  generated rewrite theories for computing $\trianglelefteq_B$ and $\trianglelefteqrogd_B$, they contain no equations, 
while the number of generated rules   increases with the complexity of the program (that heavily depends on the equational axioms that the function symbols obey). 
The number of generated rules is much bigger for $\trianglelefteqrogd_B$ than for $\trianglelefteq_B$ (for instance,   $\trianglelefteqrogd_B$ is encoded by 823 rules   
for the Dekker program versus the 59 rules of $\trianglelefteq_B$). Columns $\emptyset$,  A,C, and AC summarize the number of free, associative, commutative, 
and associative-commutative symbols, respectively, for each benchmark program. The generation times  (GT)  are negligible for all  rewrite theories.

\begin{table}[h]
	\centering
	\begin{tabular}{|l|r|r|r|r|r|r|r|r|r|r|r|r|r|}
		\hline
		\multicolumn{1}{|c|}{\multirow{2}{*}{\textbf{Benchmark}}} & \multicolumn{4}{c|}{\textbf{$\sharp$ Axioms}} & \multicolumn{3}{c|}{\textbf{$\trianglelefteq_B$}} & \multicolumn{3}{c|}{\textbf{$\trianglelefteqrogd_B$}} & \multicolumn{3}{c|}{\textbf{$\trianglelefteqdeml_B$, $\trianglelefteqdestrat_B$}} \\ \cline{2-14} 
		\multicolumn{1}{|c|}{} & \multicolumn{1}{c|}{\textbf{$\emptyset$}} & \multicolumn{1}{c|}{\textbf{~A}} & \multicolumn{1}{c|}{\textbf{~C}} & \multicolumn{1}{c|}{\textbf{AC}} & \multicolumn{1}{c|}{\textbf{~$\sharp \pr{E}$~}} & \multicolumn{1}{c|}{\textbf{~$\sharp \pr{R}$~}} & \multicolumn{1}{c|}{\textbf{GT(ms)}} & \multicolumn{1}{c|}{\textbf{~$\sharp \pr{E}$~}} & \multicolumn{1}{c|}{\textbf{~$\sharp \pr{R}$~}} & \multicolumn{1}{c|}{\textbf{GT(ms)}} & \multicolumn{1}{c|}{\textbf{~$\sharp \pr{E}$~}} & \multicolumn{1}{c|}{\textbf{~$\sharp \pr{R}$~}} & \multicolumn{1}{c|}{\textbf{GT(ms)}} \\ \hline
		\textbf{Kmp} & 9 & 0 & 0 & 0 & 0 & 15 & 1 & 0 & 57 & 2 & 21 & 0 & 0 \\ \hline
		\textbf{NatList} & 5 & 1 & 1 & 2 & 0 & 10 & 1 & 0 & 26 & 1 & 21 & 0 & 0 \\ \hline
		\textbf{Maze} & 5 & 1 & 0 & 1 & 0 & 36 & 7 & 0 & 787 & 15 & 21 & 0 & 0 \\ \hline
		\textbf{Dekker} & 16 & 1 & 0 & 2 & 0 & 59 & 8 & 0 & 823 & 18 & 21 & 0 & 0 \\ \hline
	\end{tabular}
	\vspace{1ex}
	\caption{Size of generated theories for na\"{\i}ve and goal-driven definitions vs. meta-level definitions}
	\label{tbl:sizeTheory}
\end{table}

For all benchmarks 
$T1 \trianglelefteq_B^{\alpha} T2$
in Table \ref{tbl:performance},  we have   fixed to five the size of \pr{T1}  that is measured in the depth of (the non-flattened version of) the term.
As for  \pr{T2}, we have considered terms with increasing 
depths: five, ten, one hundred, and five hundred. 
The $\sharp$ {\tt Symbols} column records the number of  A (resp.  AC) symbols occurring in the benchmarked  goals.

\begin{table}[h!]
	\centering
	\begin{tabular}{|l|c|c|r|r|r|r|r|r|}
		\hline
		\multicolumn{1}{|c|}{\multirow{2}{*}{\textbf{Benchmark}}} & \multicolumn{2}{c|}{\textbf{$\sharp$ Symbols}} & \multicolumn{2}{c|}{\textbf{Size}} & \multicolumn{1}{c|}{\textbf{$\trianglelefteq_B$}} & \multicolumn{1}{c|}{\textbf{$\trianglelefteqrogd_B$}} & \multicolumn{1}{c|}{\textbf{$\trianglelefteqdeml_B$}} & \multicolumn{1}{c|}{\textbf{$\trianglelefteqdestrat_B$}} \\ \cline{2-9} 
		\multicolumn{1}{|c|}{} & \textbf{~~A~~} & \textbf{AC} & \multicolumn{1}{c|}{\textbf{T1}} & \multicolumn{1}{c|}{\textbf{T2}} & \multicolumn{1}{c|}{\textbf{Time(ms)}} & \multicolumn{1}{c|}{\textbf{Time(ms)}} & \multicolumn{1}{c|}{\textbf{Time(ms)}} & \multicolumn{1}{c|}{\textbf{Time(ms)}} \\ \hline
		\multirow{4}{*}{Kmp} & \multirow{4}{*}{0} & \multirow{4}{*}{0} & \multicolumn{1}{r|}{\multirow{4}{*}{5}} & 5 & 10 & 6 & 1 & 1 \\ \cline{5-9} 
		&  &  & \multicolumn{1}{r|}{} & 10 & 150 & 125 & 4 & 1 \\ \cline{5-9} 
		&  &  & \multicolumn{1}{r|}{} & 100 & TO & TO & 280 & 95 \\ \cline{5-9} 
		&  &  & \multicolumn{1}{r|}{} & 500 & TO & TO & 714 & 460 \\ \hline
		\hline
		\multirow{4}{*}{NatList} & \multirow{4}{*}{1} & \multirow{4}{*}{2} & \multirow{4}{*}{5} & 5 & 2508 & 2892 & 1 & 1 \\ \cline{5-9} 
		&  &  &  & 10 & 840310 & 640540 & 1 & 1 \\ \cline{5-9} 
		&  &  &  & 100 & TO & TO & 8 & 2 \\ \cline{5-9} 
		&  &  &  & 500 & TO & TO & 60 & 5 \\ \hline
		\hline
		\multirow{4}{*}{Maze} & \multirow{4}{*}{1} & \multirow{4}{*}{1} & \multirow{4}{*}{5} & 5 & 40 & 25 & 1 & 1 \\ \cline{5-9} 
		&  &  &  & 10 & TO 
		& 20790 & 4 & 1 \\ \cline{5-9} 
		&  &  &  & 100 & TO & TO & 256 & 2 \\ \cline{5-9} 
		&  &  &  & 500 & TO & TO & 19808 & 10 \\ \hline
		\hline
		\multirow{4}{*}{Dekker} & \multirow{4}{*}{1} & \multirow{4}{*}{1} & \multicolumn{1}{r|}{\multirow{4}{*}{5}} & 5 & 50 & 40 & 1 & 1 \\ \cline{5-9} 
		&  &  & \multicolumn{1}{r|}{} & 10 & 111468 & 110517 & 2 & 1 \\ \cline{5-9} 
		&  &  & \multicolumn{1}{r|}{} & 100 & TO & TO & 5 & 3 \\ \cline{5-9} 
		&  &  & \multicolumn{1}{r|}{} & 500 & TO & TO & 20 & 13 \\ \hline
	\end{tabular}
	\vspace{1ex}
	\caption{Performance of equational homeomorphic embedding implementations w.r.t.\ problem size}
	\label{tbl:performance}

\end{table}

The figures in Table \ref{tbl:performance} confirm our expectations
regarding $\trianglelefteq_B$ and $\trianglelefteqrogd_B$ that the search space is huge and increases exponentially with the size of \pr{T2}
(discussed  for $\trianglelefteq_B$ in Example~\ref{ex:assoc-comm-pre2}
and for $\trianglelefteqrogd_B$
in Example~\ref{ex:assoc-comm}).
Actually, when  the size of \pr{T2} is 100  (and beyond) a given timeout (represented by TO in the tables)  is reached that is set for 3.6e+6 milliseconds (1 h).
The reader can also check that the more A,C, and AC symbols occur in the original program signature, the bigger the execution times.
An odd exception is the Maze example, where the timeout is already reached  for the size 10 of \pr{T2} even if the number of  equational axioms is comparable
to the other programs. 
This is because the AC-normalized, flattened version of the terms  is much smaller than the original term size for the NatList and Dekker benchmarks
but not for Maze, where the flattened and original terms have similar size.
On the other hand, our experiments demonstrate that both $\trianglelefteqdeml_B$ and $\trianglelefteqdestrat_B$ bring impressive speedups, with $\trianglelefteqdestrat_B$ 
working outstandingly well in practice even for really complex terms.

\begin{table}[h]
	\centering
	\begin{tabular}{|r|r|r|r|r|r|r|r|r|r|r|r|r|r|r|r|}
		\hline
		\multicolumn{6}{|c|}{\textbf{T1}} & \multicolumn{6}{c|}{\textbf{T2}} & \multicolumn{1}{c|}{\textbf{$\trianglelefteq_B$}} & \multicolumn{1}{c|}{\textbf{$\trianglelefteqrogd_B$}} & \multicolumn{1}{c|}{\textbf{$\trianglelefteqdeml_B$}} & \multicolumn{1}{c|}{\textbf{$\trianglelefteqdestrat_B$}} \\ \hline
		\multicolumn{2}{|c|}{\textbf{Size}} & \multicolumn{4}{c|}{\textbf{$\sharp$ Symbols}} & \multicolumn{2}{c|}{\textbf{Size}} & \multicolumn{4}{c|}{\textbf{$\sharp$ Symbols}} & \multicolumn{1}{c|}{\multirow{2}{*}{\textbf{Time(ms)}}} & \multicolumn{1}{c|}{\multirow{2}{*}{\textbf{Time(ms)}}} & \multicolumn{1}{c|}{\multirow{2}{*}{\textbf{Time(ms)}}} & \multicolumn{1}{c|}{\multirow{2}{*}{\textbf{Time(ms)}}} \\ \cline{1-12}
		\multicolumn{1}{|c|}{\textbf{OT}} & \multicolumn{1}{c|}{\textbf{FT}} & \multicolumn{1}{c|}{\textbf{~~$\emptyset$}~~} & \multicolumn{1}{c|}{\textbf{~~C~~}} & \multicolumn{1}{c|}{\textbf{~~A~~}} & \multicolumn{1}{c|}{\textbf{AC}} & \multicolumn{1}{c|}{\textbf{OT}} & \multicolumn{1}{c|}{\textbf{FT}} & \multicolumn{1}{c|}{\textbf{$\emptyset$}} & \multicolumn{1}{c|}{\textbf{C}} & \multicolumn{1}{c|}{\textbf{~~A~~}} & \multicolumn{1}{c|}{\textbf{AC}} & \multicolumn{1}{c|}{} & \multicolumn{1}{c|}{} & \multicolumn{1}{c|}{} & \multicolumn{1}{c|}{} \\ \hline
		5 & 5 & 5 & 0 & 0 & 0 & 100 & 100 & 100 & 0& 0 & 0 & 165 & 70 & 1 & 1 \\ \hline
		5 & 5 & 3 & 2 & 0 & 0 & 100 & 100 & 50 & 50& 0 & 0 & TO & 38 & 60 & 35 \\ \hline
		5 & 2 & 4 & 0& 1 & 0 & 100 & 2 &  50 & 0& 50 & 0 & TO & TO & 108035 & 3 \\ \hline
		5 & 2 & 4 & 0& 0 & 1 & 100 & 2 & 50 & 0& 0 & 50 & TO & TO & 42800 & 4 \\ \hline
		5 & 3 & 8 & 0& 1 & 2 & 100 & 3 & 50 & 0& 25 & 25 & TO & TO & 22796 & 5 \\ \hline
		5 & 5 & 5 & 0& 0 & 0 & 500 & 500 & 500 & 0& 0 & 0 & 48339 & 34000 & 12 & 4 \\ \hline
		5 & 5 & 3 & 2 & 0 & 0 & 500 & 500 & 250 & 250& 0 & 0 & TO & 2183 & 6350 & 2005 \\ \hline				
		5 & 2 & 4 & 0& 1 & 0 & 500 & 2 & 250 & 0& 250 & 0 & TO & TO & TO & 30 \\ \hline
		5 & 2 & 4 & 0& 0 & 1 & 500 & 2 & 250 & 0& 0 & 250 & TO & TO & TO & 27 \\ \hline
		5 & 3 & 8 & 0& 1 & 2 & 500 & 3 & 250 & 0& 125 & 125 & TO & TO & TO & 50 \\ \hline
	\end{tabular}
	\vspace{1ex}
	\caption{Performance of equational homeomorphic embedding implementations w.r.t.\ axiom entanglement for the NatList example}
	\label{tbl:natComparison}
\end{table}

The reader may wonder how big the impact is having A, C, or AC operators.
In order to compare the relevance of these symbols, 
in Table~\ref{tbl:natComparison}  we fix one single benchmark program (NatList) that contains all three kinds of operators: two associative operators 
(list concatenation \pr{;} and natural division \pr{/}), a commutative (natural pairing) operator (\pr{||}), and two associative-commutative arithmetic operators 
(\pr{+,*}). 
With regard to  the size of the considered terms, we confront   the size of the original term (OT) versus the size of its flattened version (FT); e.g.,\  500 versus 2 for the size of   
\pr{T2} in the last row.

We have included the execution times of $\trianglelefteq_B$ and $\trianglelefteqrogd_B$ for completeness, 
but they do not reveal a dramatic improvement of $\trianglelefteqrogd_B$ with respect to $\trianglelefteq_B$ for the benchmarked (false) goals, 
contrary to what we initially expected. This means that  $\trianglelefteqrogd_B$ cannot be generally used in real applications due to the risk of intolerable embedding test times, 
even if  $\trianglelefteqrogd_B$ may be far less wasteful than  $\trianglelefteq_B$ for succeeding goals, as discussed in Section \ref{sec:gd}. 
For $\trianglelefteqdeml_B$ and $\trianglelefteqdestrat_B$, the figures show that the more A and AC operators comparatively occur in the problem, the 
bigger the improvement achieved. 
This is due to the following:
(i) these two embedding definitions manipulate flattened meta-level terms;
(ii) they are equationally defined, which has a much better performance in Maude than doing search;
and 
(iii) our definitions are highly optimized for lists (that obey associativity) and sets (that obey both associativity and commutativity).

Homeomorphic embedding has been extensively used in Prolog for different purposes, such as  termination analysis and partial deduction. 
\begin{wrapfigure}{r}{0.5\textwidth}
	\centering
	\begin{tikzpicture}[scale=.8]
	\begin{loglogaxis}[
	xlabel={Term size}, ylabel={Time (ms)}, legend style={at={(0.05,0.85)},anchor=west}
	]
	\addplot [color=red,solid,thick,mark=*, mark options={fill=white}
	] table {
		100 302 
		500 432
		1000 1057
		5000 27938
		10000 92022
	};
	\node [above, color=red] at (axis cs:  100,  302) {302};
	\node [above, color=red] at (axis cs:  500,  432) {432};
	\node [right, color=red] at (axis cs:  1000,  1057) {1057};
	\node [left, color=red] at (axis cs:  5000,  27938) {27938};
	\node [left, color=red] at (axis cs:  10000,  92022) {92022};
	
	\addplot [color=blue,solid,thick,mark=*, mark options={fill=white}] table {
		100 15
		500 35
		1000 65
		5000 95
		10000 150
	};
	\node [above, color=blue] at (axis cs:  100,  15) {15};
	\node [below, color=blue] at (axis cs:  500,  35) {35};
	\node [below, color=blue] at (axis cs:  1000,  65) {65};
	\node [below, color=blue] at (axis cs:  5000,  95) {95};
	\node [above, color=blue] at (axis cs:  10000,  150) {150};
	
	\legend{Prolog $\trianglelefteq$ \\Maude $\trianglelefteqdestrat_\emptyset$\\}
	\end{loglogaxis}
	\end{tikzpicture}
	\vspace{-4.5ex}
	\caption{Comparison of  $\trianglelefteq$ in Prolog vs.  $\trianglelefteqdestrat_\emptyset$ for the NatList example  (no axioms in goals)}	
	\label{fig:comparisonPrologSMLl}
	\vspace{-3.5ex}
\end{wrapfigure}
In Figure~\ref{fig:comparisonPrologSMLl} we have compared on a logarithmic scale our best embedding definition,
 $\trianglelefteqdestrat_B$, 
with a standard 
meta-level Prolog\footnote{To avoid any bias, we took the Prolog code for the homeomorphic embedding of the 
	{\sc Ecce} system  \cite{LMdS99} that is available  at
	{\tt https://github.com/leuschel/ecce}, and we run it in SWI-Prolog 7.6.3.} implementation of the  (syntactic) pure homeomorphic embedding $\trianglelefteq$ of Definition \ref{def:emb-orig}. 

We chose the  NatList example and  terms \pr{T1}  and \pr{T2}  that do not contain symbols obeying equational axioms 
as this is the only case that can be handled by the    syntatic Prolog implementation.
Our experiments show that our refined deterministic 
formulation $\trianglelefteqdestrat_B$ (i.e.\ without search) outperforms the Prolog  version so  no penalty is incurred when  syntactic embeddability 
tests are run  in our equational implementation.

\label{sec:experiments}

\section{Concluding remarks}

Homeomorphic embedding has been extensively used in Prolog but it has never been  investigated in the context of expressive rule-based languages like Maude, 
CafeOBJ, OBJ, ASF+SDF, and ELAN that support symbolic reasoning methods modulo equational axioms.
We have introduced a new equational definition of homeomorphic embedding
with a remarkably good performance for theories with symbols having any combination of associativity and commutativity.
We have also compared different definitions of embedding identifying some key conclusions:
(i)~definitions of equational homeomorphic embedding based on (non-deterministic) search in Maude perform dramatically  worse than their  
equational counterparts and are not feasible in practice,
(ii)~definitions of equational homeomorphic embedding based on generated theories perform dramatically worse than 
meta-level definitions; 
and
(iii)~the flattened meta-representation of terms is crucial for   homeomorphic embedding definitions dealing with A and AC operators to pay off in practice. 
As future work, we plan to extend our results to the case when the equational theory $B$ may contain the identity axiom, which is non-trivial since $B$ is not class-finite.

\end{document}